\documentstyle{mn}
\newif\ifAMStwofonts
\def\overleftrightarrow{\mathpalette\overleftrightarrow@}
\def\overleftrightarrow@#1#2{\vbox{\ialign{##\crcr\leftrightarrowfill@#1\crcr
 \noalign{\kern-\ex@\nointerlineskip}$\m@th\hfil#1#2\hfil$\crcr}}}
\input psfig.tex

\ifoldfss
  \ifCUPmtlplainloaded \else
    \NewTextAlphabet{textbfit} {cmbxti10} {}
    \NewTextAlphabet{textbfss} {cmssbx10} {}
    \NewMathAlphabet{mathbfit} {cmbxti10} {} 
    \NewMathAlphabet{mathbfss} {cmssbx10} {} 
  \fi
  \ifAMStwofonts
    \ifCUPmtlplainloaded \else
      \NewSymbolFont{upmath} {eurm10}
      \NewSymbolFont{AMSa} {msam10}
      \NewMathSymbol{\upi}     {0}{upmath}{19}
      \NewMathSymbol{\umu}     {0}{upmath}{16}
      \NewMathSymbol{\upartial}{0}{upmath}{40}
      \NewMathSymbol{\leqslant}{3}{AMSa}{36}
      \NewMathSymbol{\geqslant}{3}{AMSa}{3E}

       \let\le=\leqslant
       \let\ge=\geqslant
    \fi
  \fi
\fi 
\ifnfssone
  \newmathalphabet{\mathit}
  \addtoversion{normal}{\mathit}{cmr}{m}{it}
  \addtoversion{bold}{\mathit}{cmr}{bx}{it}
  \newmathalphabet{\mathbfit} 
  \addtoversion{normal}{\mathbfit}{cmr}{bx}{it}
  \addtoversion{bold}{\mathbfit}{cmr}{bx}{it}
  \newmathalphabet{\mathbfss} 
  \addtoversion{normal}{\mathbfss}{cmss}{bx}{n}
  \addtoversion{bold}{\mathbfss}{cmss}{bx}{n}
  \ifAMStwofonts
    \ifCUPmtlplainloaded \else
      \UseAMStwoboldmath
      \makeatletter
      \new@mathgroup\upmath@group
      \define@mathgroup\mv@normal\upmath@group{eur}{m}{n}
      \define@mathgroup\mv@bold\upmath@group{eur}{b}{n}
      \edef\UPM{\hexnumber\upmath@group}
      \new@mathgroup\amsa@group
      \define@mathgroup\mv@normal\amsa@group{msa}{m}{n}
      \define@mathgroup\mv@bold\amsa@group{msa}{m}{n}
      \edef\AMSa{\hexnumber\amsa@group}
      \makeatother
      \mathchardef\upi="0\UPM19
      \mathchardef\umu="0\UPM16
      \mathchardef\upartial="0\UPM40
      \mathchardef\leqslant="3\AMSa36
      \mathchardef\geqslant="3\AMSa3E

       \let\le=\leqslant
       \let\ge=\geqslant
    \fi
  \fi
\fi 

\ifnfsstwo
  \DeclareMathAlphabet{\mathbfit}{OT1}{cmr}{bx}{it}
  \SetMathAlphabet\mathbfit{bold}{OT1}{cmr}{bx}{it}
  \DeclareMathAlphabet{\mathbfss}{OT1}{cmss}{bx}{n}
  \SetMathAlphabet\mathbfss{bold}{OT1}{cmss}{bx}{n}
  \ifAMStwofonts
    \ifCUPmtlplainloaded \else
      \DeclareSymbolFont{UPM}{U}{eur}{m}{n}
      \SetSymbolFont{UPM}{bold}{U}{eur}{b}{n}
      \DeclareSymbolFont{AMSa}{U}{msa}{m}{n}
      \DeclareMathSymbol{\upi}{0}{UPM}{"19}
      \DeclareMathSymbol{\umu}{0}{UPM}{"16}
      \DeclareMathSymbol{\upartial}{0}{UPM}{"40}
      \DeclareMathSymbol{\leqslant}{3}{AMSa}{"36}
      \DeclareMathSymbol{\geqslant}{3}{AMSa}{"3E}

       \let\le=\leqslant
       \let\ge=\geqslant
    \fi
  \fi
\fi 

\ifCUPmtlplainloaded \else
  \ifAMStwofonts \else 
    \def\upi{\pi}
    \def\umu{\mu}
    \def\upartial{\partial}
  \fi
\fi

\def\lsim{\lower.5ex\hbox{$\; \buildrel < \over \sim \;$}}
\def\gsim{\lower.5ex\hbox{$\; \buildrel > \over \sim \;$}}
\title{On Some Transonic Aspects of General Relativistic Spherical Accretion
onto Schwarzschild Black Holes}
\author[Tapas K. Das]
{Tapas K. Das $^{1,2}$\\
$^1$ Astronomy Unit, Queen Mary \& Westfield College, Mile End Rd, London 
E1 4NS, UK\\
e-mail: {\rm T.Das@qmw.ac.uk}\\
$^2$ Inter University Centre for Astronomy and Astrophysics
Post Bag 4 Ganeshkhind Pune 411 007 India\\
e-mail: {\rm tapas@iucaa.ernet.in}}
\date{Accepted ........  Received ......... ; in original form ........}

\begin{document}
\twocolumn
\maketitle
\begin{abstract}
The equations governing general relativistic, spherically symmetric, 
hydrodynamic accretion of polytropic fluid onto 
black holes are solved in Schwarzschild metric to investigate some of the
transonic properties of the flow. Only stationary solutions are discussed.
For such accretion, it has been shown that real physical sonic points 
may form {\it even for 
flow with $\gamma~<~\frac{4}{3}$ or $\gamma~>~\frac{5}{3}$}.
Behaviour of some flow variables in the close vicinity of the event 
horizon are studied as a function of specific energy and polytropic index 
of the flow.
\end{abstract}

\noindent
\noindent{\bf Published in the Monthly Notices of the Royal Astronomical 
Society, 2002, Volume 330, Issue 3, pp. 563-566.} \\ \\

\section{Introduction}
\noindent
Investigation of accretion processes onto celestial objects
was initiated by
Hoyle \& Littleton (1939) by computing the rate at which
pressure-less matter would be captured by a moving star. Subsequently,
theory of 
stationary, spherically symmetric and transonic hydrodynamic accretion of
adiabatic fluid onto a gravitating astrophysical object at rest was
formulated in a seminal paper by Bondi (1952) using purely Newtonian potential
and by including the pressure effect of the accreting material.
Later 
on, Michel (1972) discussed fully general relativistic polytropic accretion on
to a Schwarzschild black hole by formulating the governing equations for steady
spherical flow of perfect fluid in Schwarzschild metric. Following
Michel's relativistic generalization of Bondi's treatment,
Begelman (1978) discussed some aspects of the critical points of the 
flow for such an accretion. 
Spherical accretion and wind in general relativity have also been considered
using equations of state other than the polytropic one and
by incorporating various radiative processes 
(Shapiro, 1973a,b, Blumenthal \& Mathews 1976,
Brinkmann 1980).
Recently Malec (1999) provided 
the solution for general relativistic  spherical accretion with and 
without back reaction and showed that relativistic effects enhance mass 
accretion when back reaction  is neglected.\\
\noindent 
It is to be noted that one very important issue in understanding 
the flow profile for accretion onto gravitating astrophysical 
objects is the `transonicity' of the flow. 
Let the instantaneous dynamical velocity and the local acoustic velocity
of a compressible fluid moving along a space curve parameterized by $r$
be $u(r)$ and $a(r)$ respectively. Local Mach number $M(r)$ of the
fluid can then be defined as the ratio of the dynamical flow velocity to
its sound speed, i.e., $M(r)=\left[\frac{u(r)}{a(r)}\right]$.
The flow will be locally subsonic or supersonic according to $M(r) < 1$ 
or $ >1$, i.e., according to
$u(r)<a(r)$ or $u(r)>a(r)$. The flow is transonic if at any moment 
it crosses $M=1$. This happens when subsonic to supersonic or supersonic to
subsonic transition takes place either continuously or discontinuously.
The points where such crossing continuously 
takes place are called sonic points
and where such crossing takes place discontinuously are called shocks
or discontinuities.
One crucial difference between the flow characteristics around a black hole and 
around any other astrophysical object is, as the flow approaches to any 
type of accretor other than a black hole, it can `physically' hit the 
surface of the accretor directly and at the particular moment it 
collides with the surface, accretion can be either subsonic or supersonic
depending on the location of the sonic point (which itself is a function
of various accretion parameters) as well as on the location of the 
surface of the accretor whereas for accretion onto black holes, 
flow must fall onto the black hole 
only supersonically because even for the steepest possible equation of state,
the maximum possible sound velocity attained will always be lower than the 
bulk velocity of the flow at the event horizon.
On the otherhand,
it is quite possible that the flow at a sufficiently large distance away
from the accretor, would be subsonic in general (except for a special case 
when the supersonic wind from nearby object(s) falls onto the accretor). So
black hole accretion is {\it necessarily transonic} to satisfy the 
inner boundary condition at the event horizon whereas accretion onto other
class of astrophysical objects may not always be transonic.\\
\noindent
One standard method to study the 
classical transonic Bondi flow is to formulate the basic conservation 
equations for the flow and then to 
simultaneously  solve these conservation equations to get
sonic  quantities as function of various accretion parameters
and also to calculate the values of various dynamical as well as thermodynamic
quantities as functions of various accretion
parameters or radial distance (measured from the central
accretor in the unit of Schwarzschild radius).
However, the most effective approach to study the transonicity of 
spherically symmetric black hole accretion, as we believe, is to 
investigate the dependence of the location of the flow sonic point
on various accretion parameters as well as to study the variation
of Mach number of the flow with radial distance; which, perhaps, has
not been explicitly discussed in existing literature
for full general relativistic description of accretion
onto black hole. In this paper, we would like to
address this issue by formulating and solving the required equations 
of motion for a spherically symmetric transonic polytropic fluid accretion
in Schwarzschild metric.
\section{Governing Equations}
\noindent
We take the Schwarzschild radius $r_g$ to be:
$$
r_g=\frac{2G{M_{BH}}}{c^2}
$$ 
where  $M_{BH}$  is the mass of the black hole, $G$ 
is universal gravitational  
constant and $c$ is velocity of light in vacuum. 
We assume that a Schwarzschild type black hole 
spherically accretes fluid
obeying polytropic equation of state. The density of the fluid is $\rho(r)$,
$r$ being the radial distance measured in the
unit of Schwarzschild radius $r_g$. We also assume that the accretion rate
is not a function
of $r$ and we ignore the
self-gravity of the flow.
For simplicity of calculation, we choose gravitational unit
where unit of length is scaled in units of $r_g$, unit of velocity is  
scaled in units of $c$ and all other physically relevant
quantities can be normalized likewise. We also set $ G=c=1$ in system of
units used here.\\
\noindent
For a Schwarzschild metric of the form  
$$
ds^2=dt^2\left(1-\frac{1}{r}\right)-dr^2{\left(1-\frac{1}{r}\right)}^
{-1}-r^2{\left(d{\theta}^2+sin^2{\theta}{d\phi^2}\right)}
$$
the energy momentum tensor $T^{{\alpha}{\beta}}$ for a perfect fluid
can be written as (Shapiro \& Teukolsky 1983):
$$
T^{\alpha\beta}={\epsilon}u^{\alpha}u^{\beta}+p\left(u^{\alpha}u^{\beta}-
g^{\alpha\beta}\right)
$$
where ${\epsilon}$ and $p$ are proper internal 
energy density and pressure of the
fluid (evaluated in the local inertial rest frame of the fluid)
respectively and $u^{\alpha}$ is the four velocity commonly known as
$$
u^{\alpha}=\frac{dx^{\alpha}}{ds}
$$
Equations of motion which are to be solved for our purpose are,\\
1) Conservation of mass flux or baryon number conservation:
$$
{\left({\rho}{u_{\alpha}}\right)}_{;~{\alpha}}=0
\eqno{(1a)}
$$
and \\
2) Conservation of momentum or energy flux (general relativistic Euler
equation obtained by taking the four divergence of $T^{{\alpha}{\beta}}$):
$$
\left({\epsilon}+p\right){u_{{\alpha};{\beta}}}u^{\beta}=
-p_{,\alpha}-u_{\alpha}p_{,\beta}u^\beta_{,}
\eqno{(1b)}
$$
where $\rho$ is the proper matter density and
the semicolons denote the covariant derivatives. \\
\noindent
Following Michel (1972), one can rewrite eq. 1(a) and eq. 1(b) for
spherical accretion as
$$
4{\pi}{\rho}ur^2={\dot M}_{in}
\eqno{(2a)}
$$
and
$$
\left(\frac{p+\epsilon}{\rho}\right)^2\left(1-\frac{1}{r}+u^2\right)={\bf C}
\eqno{(2b)}
$$
as two fundamental conservation equations for time
independent hydrodynamical flow of matter on to a
Schwarzschild black hole without back-reaction of the
flow on to the metric itself. ${\dot M}_{in}$   being the mass
accretion rate and ${\bf C}$ is some constant
(related to the total enthalpy influx)
to be evaluated
for a specific equation of state.\\
\noindent
It is well known in general theory of relativity that a stationary
and axisymmetric space-time is endowed with one space-like 
$\left[\left(\frac{\partial}{{\partial}{\phi}}]\right)^{\alpha}\right]$
and one time-like 
$\left[\left(\frac{\partial}{{\partial}{t}}]\right)^{\alpha}\right]$
Killing field where $\phi$ is the standard azimuthal co-ordinate. 
For a simpler space-time which is spherically symmetric,
only the time like Killing filed is of particular interest and corresponding
to this field, one can obtain the integral of motion along a streamline as:
$$
{\cal E}=h\epsilon_b=\frac{p+\epsilon}{\rho}\epsilon_b
\eqno{(3)}
$$
where ${\cal E}$, $h\left(=\frac{p+\epsilon}{\rho}\right)$ and
$\epsilon_b$ are the conserved specific energy of the flow 
including its rest mass, specific enthalpy and specific binding 
energy respectively. For polytropic equation of state
$p=K{\rho}^{\gamma}$ (where $\gamma$ is the polytropic index and
$K$ is a constant which can be 
considered as the measure of the entropy of the flow), the generalized
expression for the sound velocity $a=\left(\frac{{\partial}p}
{{\partial}\epsilon}\right)_{\cal S}^{\frac{1}{2}}$ gives (Frank et. al. 1992,
Weinberg 1972):
$$
a=\sqrt{\frac{{\gamma}{p}}{\rho}}=
\sqrt{\frac{{\gamma}{\kappa}{T}}{{\mu}{m_H}}}
\eqno{(4)}
$$
where $T$ is the flow temperature, $\mu$ is the mean molecular weight and
$m_H{\sim}m_p$ is the mass of the Hydrogen atom. 
The subscript ${\cal S}$ indicates that
differentiation is performed at constant specific entropy.
Also one can easily show 
that the specific enthalpy $h$ of the flow is related to the sound speed
through the following equation
$$
h=\left(1-\frac{a^2}{\gamma-1}\right)^{-1}
\eqno{(5)}
$$
\\
\noindent
Using the above expression for enthalpy,
one can easily rewrite 
the conservation equation (2b) as the specific energy conservation as
$$
{\cal E}=\left[\frac{\gamma-1}{\gamma-\left(1+a^2\right)}\right]
\left(\frac{1-\frac{1}{r}}{1-u^2}\right)^{\frac{1}{2}}
\eqno{(6a)}
$$
\noindent
Defining ${\dot {\cal M}}={\dot M}_{in}{\gamma}^{\frac{1}{\gamma-1}}
K^{\frac{1}{\gamma-1}}$ to be another constant of motion for a shock
free polytropic flow, one can rewrite eqn. (2a) as (see Chakrabarti (1996)
and references therein):\\
$$
{\dot {\cal M}}=4{\pi}\left[\frac{\left(\gamma-1\right)a^2}
{\gamma -\left(a^2+1\right)}\right]
^{\left(\frac{1}{\gamma-1}\right)}
u{\epsilon_b}r^2
\eqno{(6b)}
$$
The above equation may be considered as the outcome of the conservation of mass 
and entropy along the flow line.\\
\noindent
One can now easily derive the expression for velocity
gradient $\left(\frac{du}{dr}\right)$ (by differentiating eq. 6(a) and 6(b))
as 
$$
\frac{du}{dr}=\frac{u\left(1-u^2\right)\left[a^2\left(4r-3\right)-1\right]}
{2r\left(r-1\right)\left(u^2-a^2\right)}
\eqno{(7a)}
$$
Since the flow is assumed to be smooth everywhere, if
the denominator of eq. 7(a)  vanishes at any radial distance
$r$, the numerator must also vanish there to maintain the
continuity of the flow. One therefore arrives at the so
called `sonic point  conditions'  by simultaneously making
numerator and denominator of eq. 7(a) equal to zero and
the sonic point conditions can be expressed as follows
$$
u_c=a_c=\sqrt{\frac{1}{4r_c-3}}
\eqno{(7b)}
$$
here suffix $c$ indicates that the values of the respective quantities are 
measured at the sonic point of the flow.
For a specific value of ${\cal E}$ and $\gamma$,
location of sonic point $r_c$
can be obtained by solving the following equation
algebraically for $r_c$
$$
64r_c^3\left({\cal E}^2-1\right)-
16r_c^2\left(2{\cal E}^2{\psi}-9\right)
+4r_c\left({\cal E}^2{\psi^2}-27\right)+27=0
\eqno{(7c)}
$$
where ${\psi}=\left(\frac{3\gamma+2}{\gamma-1}\right)$.
It is important to note that though eqn. 7(c) is a third order polynomial
in $r_c$, for all values of ${\cal E}$ and ${\gamma}$, it  gives
only one real physical root for $r_c$ in general,
we will return to this issue in next section. \\
\noindent
The spherical surface of radius $r=r_c$ can be defined as
`acoustic horizon' because for $r~<~r_c,~u~>~a$;  and any
acoustic disturbances created in this region are
advected towards the black hole. Thus no acoustic
disturbances created within this radius can cross the
acoustic horizon and escape to the region $r~>~r_c$. \\
\noindent
To
determine the behaviour of the solution near the sonic
point, one needs to evaluate the value of $\left(\frac{du}{dr}\right)$
at that
point (we denote it by $\left(\frac{du}{dr}\right)_c$)
by applying L `Hospitals' rule on eq. 7(a). It is
easy to show that $\left(\frac{du}{dr}\right)_c$
can be obtained by solving the
following quadratic equation algebraically:
$$
\left(\frac{du}{dr}\right)^2_c+\frac{\left(\gamma-1\right)
\left(16r^2_c-16r_c-8{\gamma}r_c+6\gamma+3\right)}
{3r_c\left(4r_c-3\right)^{\frac{3}{2}}}
\left(\frac{du}{dr}\right)_c
$$
$$
+
\frac{\left(\gamma-1\right)\left(2r_c-1\right)\left(24r^2_c-28r_c-8r^2_c\gamma+4
r_c\gamma+3\gamma+6\right)}
{2r_c\left(4r_c-3\right)^{\frac{3}{2}}\left(4r_c-3\right)\left(r_c-1\right)}
=0
\eqno{(7d)}
$$
It is now quite straightforward to simultaneously
solve eq. 6(a) and eq. 6(b) to
get the integral curves of the flow 
(curves showing the variation of Mach number with radial distance)
for a fixed value
of ${\cal E}$ and $\gamma$. Detail methodology for this purpose will 
be discussed
in \S 3.\\
\begin{figure}
\vbox{
\vskip -1.0cm
\centerline{ 
\psfig{file=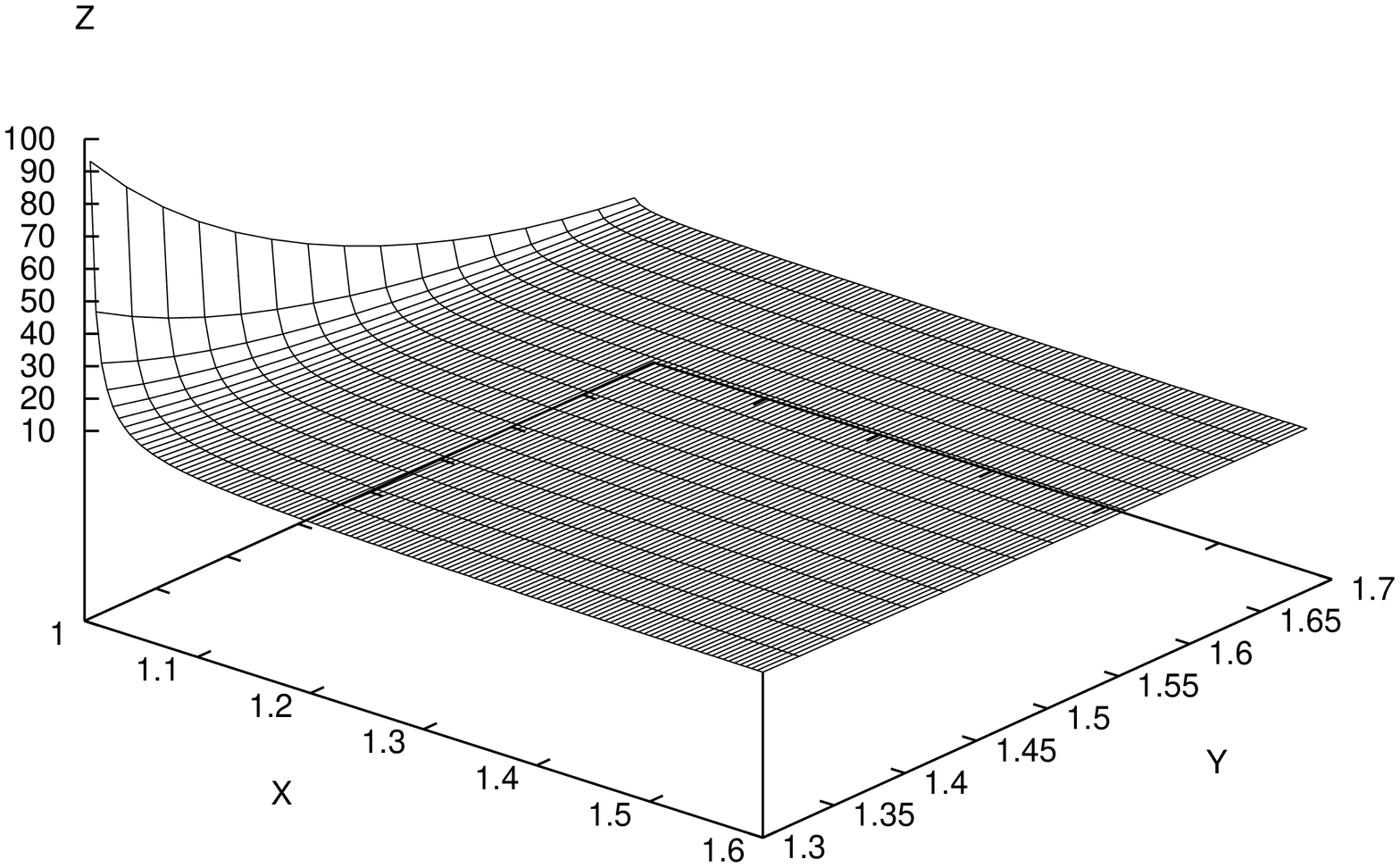,height=10cm,width=10cm,angle=270.0}}} 
\noindent {{\bf Fig. 1:} 
Variation of the location of the accretion sonic point $r_c$ with
conserved specific energy ${\cal E}$ (which includes the rest mass
energy of matter) and polytropic index $\gamma$ of the flow. While 
$r_c$ is plotted along z axis, ${\cal E}$ and $\gamma$ are plotted along 
x and y axis respectively. 
The
curves shown in the figure are plotted corresponding  to the
increments of 0.005 
and 0.025
in ${\cal E}$ and $\gamma$ respectively.
Note that $r_c$ is produced even for flow 
with $\gamma\le4/3$ and $\gamma\ge5/3$, though not all sonic points
shown here allow real physical transonic accretion. See text for detail.}
\end{figure}
\section{Solution Procedure and Results}
\noindent
One can obtain the location of the flow sonic point by
solving eqn. 7(c) for a fixed value of the conserved specific energy
${\cal E}$ and polytropic index $\gamma$ of the flow. As already mentioned,
only one physical value of $r_c$ would be obtained which lies outside
the event horizon. We solve eqn 7(c) for a range of values of ${\cal E}$
and $\gamma$ for which a real physical solution is possible. There are 
cases where only one real solution for $r_c$ is present but it lies
inside the event horizon so we exclude those solutions. In Fig. 1., we
show the variation of $r_c$ (plotted along z axis) with ${\cal E}$ 
(plotted along x axis) and 
$\gamma$ (plotted along y axis). It is observed that the location of the sonic
point non-linearly anti-correlates with both ${\cal E}$ and $\gamma$ which 
implies that ultra-relativistic flow 
\footnote{Hereafter, we will describe the flow to be ultra-relativistic
for $\gamma=\frac{4}{3}$ and purely non-relativistic for $\gamma=\frac{5}{3}$
according to standard practice (Frank. et. al 1992).}
with low total specific energy will produce the sonic point located 
furthest distance away from the black hole.
One thing is very interesting in Fig. 1. We observe that sonic point is 
produced even for flows with ${\gamma \le 4/3}$ and ${\gamma \ge 5/3}$
though not all such sonic points allow steady physical transonic flows
passing through them, which would be clear from the following discussion and from
Fig. 3.
\begin{figure}
\vbox{
\vskip -3.5cm
\centerline{ 
\psfig{file=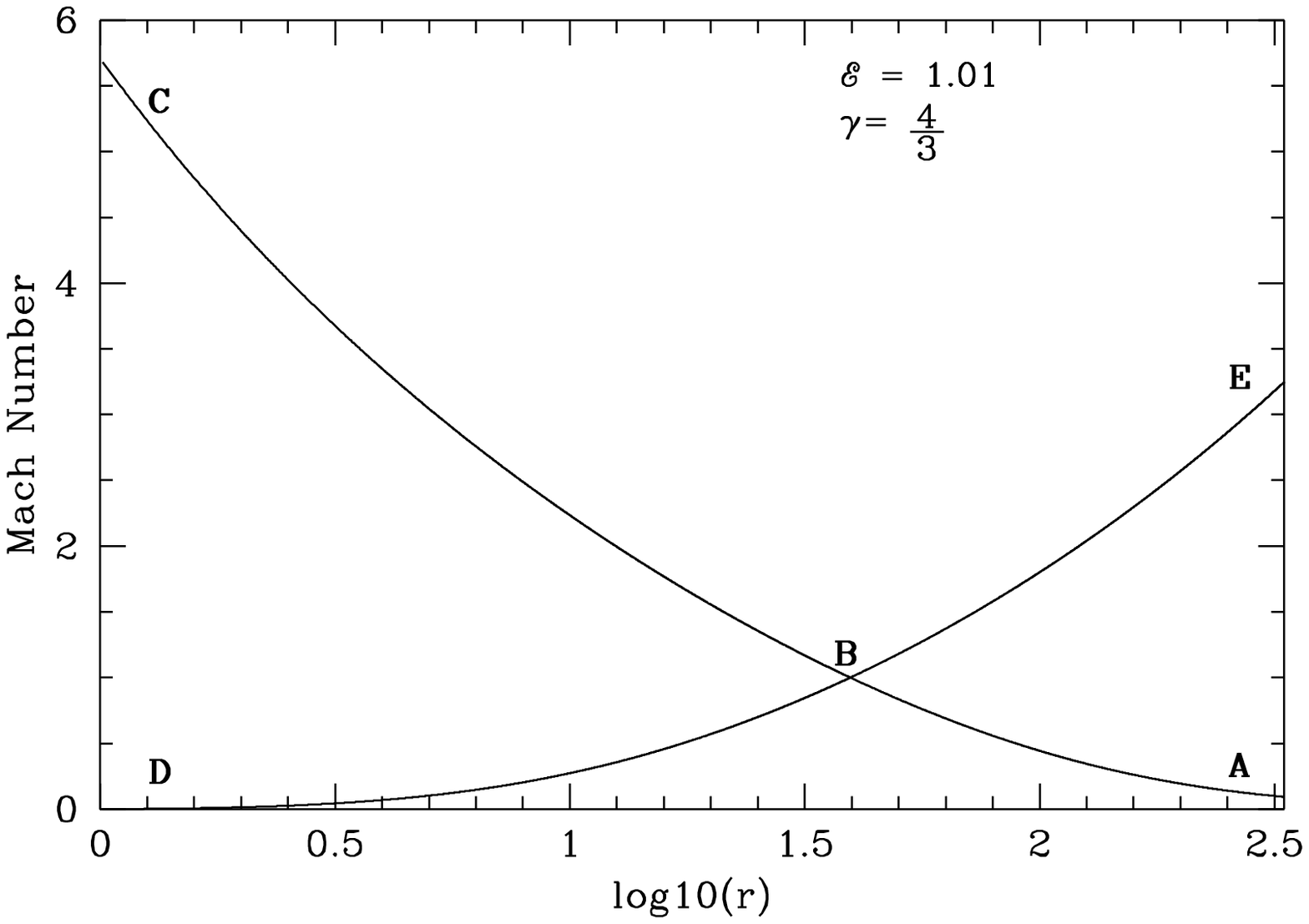,height=12cm,width=10cm,angle=0.0}}} 
\noindent {{\bf Fig. 2:} 
Integral curves of motion for a fixed value of ${\cal E}=1.01$ and
$\gamma=4/3/$. While Mach number of the flow M is plotted along y axis,
distance from the central accretor (in units of $r_g$) is plotted along
x axis in logarithmic scale.}
\end{figure}

\noindent
As we have mentioned above, $r_c$ can be obtained from eqn. 7(c). 
Now the value of $\left(\frac{du}{dr}\right)$ at $r_c$ can be obtained 
by solving eqn. 7(d) for  a particular value of ${\cal E}$ and $\gamma$. 
Fourth order Runge-Kutta method is then employed to integrate from sonic
point to get the dynamical flow velocity and the acoustic velocity (so also
the Mach number of the flow) at any point of the flow. It is well
known that for spherically symmetric accretion in Newtonian potential, two
solutions are obtained while solving the conservation equations, one out 
of which corresponds to the accretion and the other one to the `self-wind'.
Same kind of situation is obtained here also for flows in general relativity.
In Fig. 2 we plot variation of Mach number (plotted along y axis) with 
radial distance from the accretor (along x axis in unit of $r_g$) in 
logarithmic scale for accretion (ABC) and wind (DBE) branches for a fixed 
value of ${\cal E}$ (=1.01) and $\gamma$ (=4/3). The location of the sonic point
B comes out to be $39.53~r_g$. It is important to note that though from eqn 7(c),
sonic points are obtained for a wide range of values of ${\cal E}$ and $\gamma$,
in reality, not all sonic points allow a real physical flow to pass through them.
We find that for accretion passing through some of the sonic points 
(obtained for a specific region of parameter space spanned by ${\cal E}$
and $\gamma$), dynamical velocity or acoustic velocity, or both, becomes 
superluminal hence causality relation is violated which indicates that
only a subset of the sonic points shown in Fig. 1 can be considered to 
study the transonicity of a real physical accretion flow. Here we would like to 
emphasize that still some sonic points, which allow real physical flow passing 
through them, are obtained for accretion with $\gamma \le \frac{4}{3}$
or $\gamma \ge \frac{5}{3}$, see Fig. 3.\\
\noindent
We would now like to investigate the behaviour of flow variables close to the 
inner boundary of accretion. For this purpose we would like to study the the Mach number of the 
flow and the flow temperature very close to the event horizon of the 
black hole. As all equations diverge on the event horizon, we would like to
tackle the problem in the following way:\\
\noindent
Let $r_h$ be the actual location of the event horizon (in units used here,
$r_h=r_g$) and let $\Delta r$ be equal to ${\delta}r_g$ where $\delta$ is a small
number less than one. We then define $r^{ex}=r_h+\Delta r$ to be the extreme 
point for our calculation of any flow variable 
\begin{figure}
\vbox{
\vskip -3.5cm
\centerline{
\psfig{file=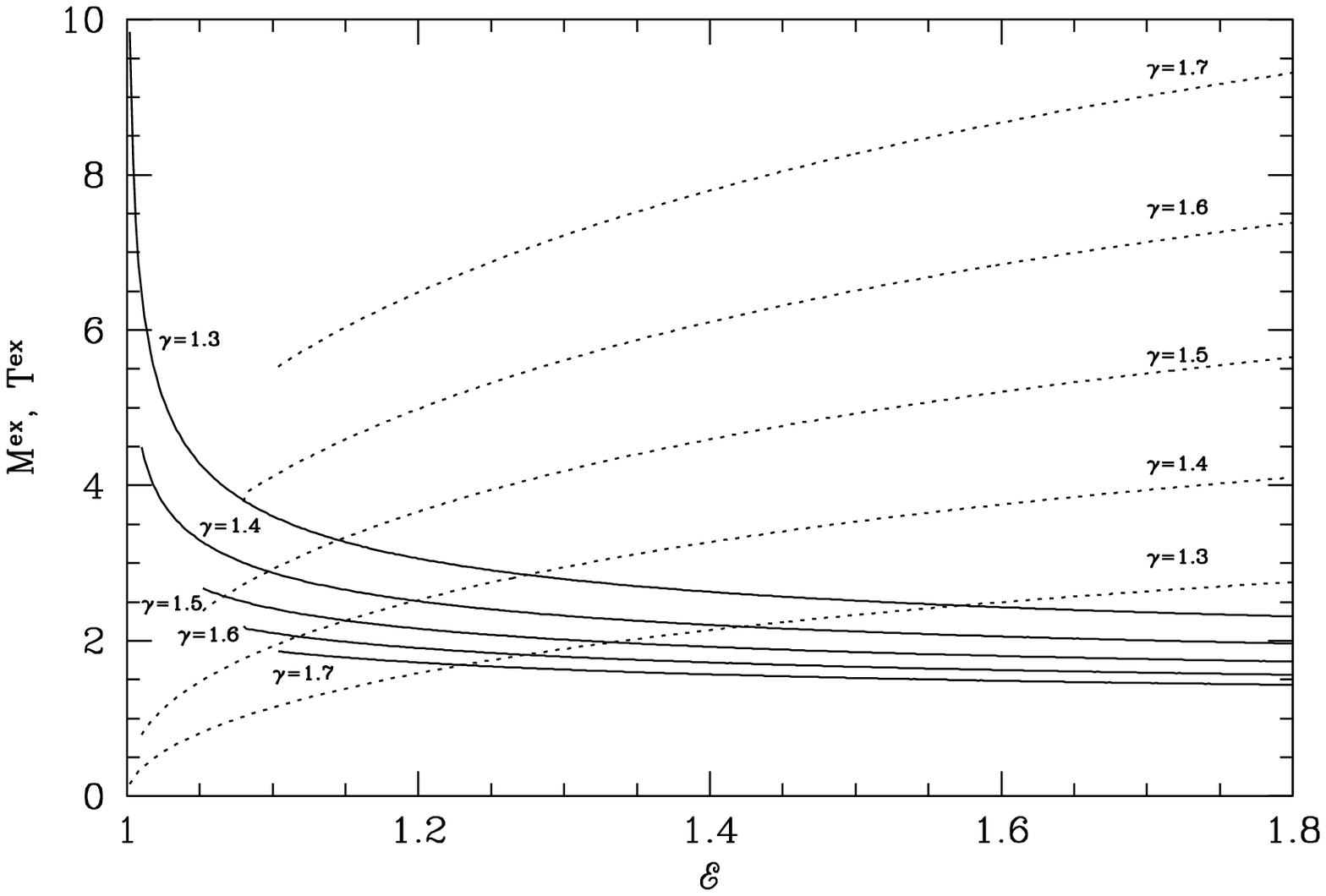,height=12cm,width=10cm,angle=0.0}}}
\noindent {{\bf Fig. 3:}
Behaviour of Mach number of the flow $M$ (solid lines) and the flow temperature
$T$
(dotted lines) for a set of values of ${\cal E}$ and $\gamma$ marked in the
figure. $M^{ex}$ and $T^{ex}$ for the values of $M$ and $T$ respectively
measured at a distance
$0.01r_g$ away from the event horizon.  $T^{ex}$ is scaled as
$T^{ex}\longrightarrow 0.22~T^{ex}/T_{11}$ (where $T_{11}=10^{11}~{^o\!K}$)
to fit in the same graph; see text for details.}
\end{figure}
$V_{flow}$ so that $V_{flow}^{ex}$
refers to the value of $V_{flow}$ measured at a radial distance 
($1+\delta$) $r_g$. Here we would like to calculate $M^{ex}$ and 
$T^{ex}$ for $\delta=0.01$ for a fixed values of ${\cal E}$ and $\gamma$ and to
study the variation of $M^{ex}$ and $T^{ex}$ with ${\cal E}$ and $\gamma$.
Lower value of $\delta$ can also be taken but it would only increase the 
computational cost, the general solution profile would remain unaffected.\\
\noindent
In what follows, for a particular value of  ${\cal E}$ and $\gamma$, we solve
eqn 6(a) and 6(b) simultaneously upto $r^{ex}$ to get $M^{ex}$ and
in the same way we calculate $M^{ex}$ for various values of 
${\cal E}$ and $\gamma$ which allows real physical solutions. Similarly,
as the sound speed $a$ can be calculated easily while calculating $M$, one can
obtain $T^{ex}$ (using eqn. (4))
for various values of ${\cal E}$ and $\gamma$. In Fig. 3 we show the variation
of $M^{ex}$ (solid line) and $T^{ex}$ (dotted line) with ${\cal E}$ for a set of
equi-spaced values of $\gamma$ shown in the figure. $T^{ex}$ has been normalized
as $T^{ex}\longrightarrow 0.22~T^{ex}/T_{11}$ (where $T_{11}=10^{11}~{^o\!K}$)
to fit in the same graph. 
One can note that the value of 
minimum energy ${\cal E}_{min}$ for which real physical accretion
solution is available, non-linearly increases with increase of $\gamma$.
We see that $M^{ex}$ non-linearly 
anti-correlates with ${\cal E}$ and ${\gamma}$. This is obvious because as we have 
seen from Fig. 1, flow with 
low energy and low $\gamma$ produces the sonic 
point far away from the black hole so that the flow becomes more and more
supersonic as it approaches the event horizon.
However, $T^{ex}$ follows exactly the opposite profile, it non-linearly 
correlates with both ${\cal E}$ and $\gamma$ because for high energy accretion
as well as for high enthalpy accretion (which is equivalent to high $\gamma$ 
flow), flow temperature becomes higher. It is important to note that 
for a lower value of $\delta$, both $M^{ex}$ and $T^{ex}$ keeps
increasing keeping the general $\left(M^{ex},T^{ex}\right)$
vs $\left({\cal E},\gamma\right)$ profile remain unaltered.
\section{Conclusion}
\noindent
In this paper we have solved the equations governing general relativistic
spherically symmetric hydrodynamic accretion of polytropic fluid on to 
a Schwarzschild black hole to investigate some of the transonic properties
of the flow. We could study the variation of flow critical points with the 
conserved specific energy ${\cal E}$ and the polytropic index $\gamma$ of the 
flow and showed that while high energy purely-non-relativistic accretion
produces sonic points closer to the event horizon of the black hole, 
low energy ultra-relativistic flows pushes the sonic point far away from the 
hole. Also it is shown that not all sonic points allows a real physical flow
to pass through them. Perhaps the most important
finding in this paper regarding the transonicity of the flow
is that it is possible to obtain real physical sonic points 
(which allows a real physical general relativistic spherically symmetric
transonic accretion through them) even for accretion with 
$\gamma~<~\frac{4}{3}$ or $\gamma~>~\frac{5}{3}$.
Also we study the behaviour of the flow temperature
and Mach number of the flow in close vicinity of the event horizon of the 
black hole as a function of ${\cal E}$ and $\gamma$ and show that while low
energy ultra-relativistic flow produces high Mach number at event horizon, high 
energy purely-non-relativistic flow produces high temperature at the horizon.
\section{Acknowledgments}
The author would like to acknowledge the hospitality provided by the Astronomy 
Unit,
School of Mathematical Sciences, Queen Mary \& Westfield College, University of 
London, 
where a part of this research was carried out.
\noindent


\begin{thebibliography}{}
\bibitem[]{} Begelman, M. C. 1978, A \& A, 70, 583
\bibitem[]{} Blumenthal, G. R., \& Mathews, W. G. 1976, ApJ, 203, 714
\bibitem[]{} Bondi, H. 1952, MNRAS, 112, 195
\bibitem[]{} Brinkmann, W. 1980, A \& A, 85, 146
\bibitem[]{} Chakrabarti, S. K. 1996, Phys. Reports., 266, No 5 \& 6, 229
\bibitem[]{} Frank, J., King, A., \& Raine, D. 1992, Accretion Power in
Astrophysics. 2nd. Edition. Cambridge University Press.
\bibitem []{} Hoyle, F., \& Lyttleton, R. A. 1939, 
Proc. Camb. Phil. Soc, 35, 592
\bibitem[]{}Malec, E. 1999, Phys. Rev. D. 60, 104043
\bibitem[]{} Michel, F. C. 1972, Astrophys. Space Sci. 15, 153
\bibitem[]{} Shapiro, S. 1973a, ApJ, 180, 531
\bibitem[]{} Shapiro, S. 1973b, ApJ, 185, 69
\bibitem[]{} Shapiro, S. L., \& Teukolsky, S. A. 1983, Black Holes,
White Dwarfs and Neutron stars. John Wiley and Sons Inc.
\bibitem []{} Weinberg, S. 1972, Gravitation and Cosmology : Principles and
Applications
of the General Theory of Relativity (John Wiley \& Sons)
\end{thebibliography}
\end{document}